

\font\titolino=cmbx10
\font\tsnorm=cmr10
\font\tscors=cmti10

\font\tscorsp=cmti9
\magnification=1200
\hsize=148truemm
\hoffset=10truemm
\parskip 3truemm plus 1truemm minus 1truemm
\parindent 8truemm
\newcount\notenumber

\def\PR{{\tscors Phys. Rev. }}
\def\NP{{\tscors Nucl. Phys. }}

\def\MPL{{\tscors Mod. Phys. Lett. }}

\def\CQG{{\tscors Class. Quantum Grav. }}
\def\note{\advance\notenumber by 1 \footnote{$^{\the\notenumber}$}}
\def\ref#1{\medskip\everypar={\hangindent 2\parindent}#1}
\def\beginref{\begingroup
\bigskip
\leftline{\titolino References.}
\nobreak\noindent}
\def\endref{\par\endgroup}
\def\beginsection #1. #2.
{\bigskip
\leftline{\titolino #1. #2.}
\nobreak\noindent}
\def\beginappendix #1.
{\bigskip
\leftline{\titolino Appendix #1.}
\nobreak\noindent}

\nopagenumbers
\rightline{}
\vskip 20truemm
\centerline{\titolino CAN THE INTERACTION BETWEEN BABY UNIVERSES}
\bigskip
\centerline{\titolino GENERATE A BIG UNIVERSE?}
\vskip 15truemm
\centerline{\tsnorm Marco Cavagli\`a}
\bigskip
\centerline{\tscorsp ISAS, International School for Advanced
Studies, Trieste, Italy}
\centerline{\tscorsp and}
\centerline{\tscorsp INFN, Sezione di Torino, Italy.}
\vfill
\centerline{\tsnorm ABSTRACT}
\begingroup\tsnorm\noindent
We explore a simple toy model of interacting universes to establish that
a small baby universe could become large ($\gg$ Planck length) if a third
quantization mechanism is taken into account.
\vfill
\centerline{Ref. SISSA 104/94/A}
\vfill
\hrule
\noindent
Mail Address:
\hfill\break
ISAS-International School for Advanced Studies
\hfill\break
Via Beirut 2-4, I-34013 Miramare (Trieste)
\hfill\break
Electronic mail: 38028::CAVAGLIA or CAVAGLIA@TSMI19.SISSA.IT
\endgroup
\eject
\footline{\hfill\folio\hfill}
\pageno=1
\noindent
In quantum cosmology the Wheeler--de Witt (WDW) equation describes the
dynamics of a single universe as a whole. Many solutions of the WDW
equation that can be interpreted as wave functions of the universe are
known in the literature but all seems unsuccessful in describing our
universe; as in the classical framework, also in the quantum theory a
cosmological constant causing the inflation of the Planck-scale baby
universe solution of the WDW equation is required to explain the main
properties of our universe (flatness, oldness...): in minisuperspace
models this fact can be easily understood observing how the presence of a
cosmological term modifies the harmonic potential in the WDW equation and
allows the tunnelling from ``nothing'' to a de Sitter universe (see,
for instance, [1]). In this paper we will establish how a solution
describing a universe of size $\gg l_{Pl}$ becomes natural if the
interaction with a quantum baby universes spacetime foam is taken into
account. In the following we will discuss a simple toy model and hope
that this paper will produce further discussions on this argument.

As we have mentioned, the WDW equation ($H\psi=0$) describes the
dynamics of a single universe [2,3] and it is essentially a zero
energy Schr\"odinger equation in the superspace. Generally we can
assume that a complete set of solutions do exist: we label this set
by an index $k$. To discuss interactions between universes, we assume
the existence of a field theory on superspace whose free field
equation is given by the WDW equation. Therefore we form linear
combinations of the creation and destruction operators of universes
$c^\dagger$ and $c$:
$$\eqalign{&\Psi(h)=\sum_k\psi_k(h) c_k,\cr
&\Psi^\dagger(h)=\sum_k\psi^\dagger_k(h) c^\dagger_k,\cr}\eqno(1)$$
where $h$ represents the three-dimensional metric of the manifold
mapping the superspace. $\Psi$ and $\Psi^\dagger$ are field operators in
the abstract occupation number Hilbert space and the operators $c_k$,
$c^\dagger_k$ satisfy the boson commutation relations (the universes are
bosons). The kinetic term of the field theory action takes the form
[4,5]:
$$S=-{1\over 2}\int Dh\Psi^\dagger H\Psi.\eqno(2)$$
Then it is not difficult to convince itself that the interactions
between universes modify the WDW equation (tree level) with a
potential term [4]:
$$H\Psi=-{dV[\Psi]\over d\Psi}.\eqno(3)$$
The idea is that an observer in a given universe could interpret the
potential term as an effective term in the WDW equation of the
universe where he lives. In this way it is possible that the wave
function modified by the interactions could describe a universe
similar to ours in a natural way without the presence of a
cosmological constant. To construct our model we suppose to work in
the minisuperspace, i.e. we consider only spatially homogeneus and
isotropic closed universes; in this case the WDW equation reduces to
a one-dimensional zero energy Schr\"odinger equation in the scale
factor variable $a$. We suppose also that the matter contribution to
the WDW equation separates from the gravitational contribution: this
is not a very strange requirement because the conformal scalar field
or the Yang-Mills radiation field satisfy this condition [6-8]. With
these assumptions the WDW equation separates in the gravitational
($H_g(a)$) and matter ($H_M(\phi)$) parts and the wave function can be written
(in the following we put the Planck length $l_{Pl}=1$):
$$\psi_k(a,\phi)=\chi_k(a)\Phi(\phi,k)\eqno(4)$$
where $\Phi$ represents the matter part of the wave function and
$\chi_k$ is the $k$-th harmonic oscillator wave function
($H_g\chi_k=\epsilon_k\chi_k$); the energy $\epsilon_k$
is related to the energy density of the matter field $\phi$. Then (4)
represents a quantum closed spherically symmetric universe and for
large values of the energy, i.e. for large quantum numbers ($k\approx
10^{120}$), (4) describes our universe [2,8]: as in the classical
theory our universe appears very strange! In fact a typical solution
(4) has small quantum numbers, i.e. dimensions of order of the Planck
length. We suppose now the existence of a foam of baby universes (4)
and a two body interaction $v(a,a')$ between universes in the
minisuperspace.  The gravitational part of the single
universe quantum equation in the Hartree--Fock [9] approximation can
be written:
$$\bigl(H_g+V_H\bigr)\tilde\chi_k=\tilde\epsilon_k\tilde\chi_k\eqno(5)$$
where $V_H$ is the Hartree--Fock potential
$$V_H=\int da'\sum_l\tilde\chi_l^2(a')v(a,a').\eqno(6)$$
In particular if we consider $N-1$ universes in the foam ground
state ($k=0$) and one universe in an excited state with a
small quantum number, the one-body equation for the excited universe
becomes ($v(a,a')=-gv(a)\delta(a-a')$, where $g$ is a positive
coupling constant $O(1)$):
$${1\over 2}\biggl[-{d^2\over da^2}+a^2-
2g(N-1)v(a)\tilde\chi_0^2\biggr]\tilde\chi_k=
\tilde\epsilon_k\tilde\chi_k.\eqno(7)$$
Choosing a potential $v(a)\approx\exp(a^2)$, (5) can be cast in the form
of a Schr\"odinger equation for a harmonic oscillator. This potential can
really arise when two universes are involved [6], being $\exp(-S)$ the
generic process amplitude; the energy levels are:
$$\tilde E_k\approx\tilde\epsilon_k+g(N-1)=(k+1/2).\eqno(8)$$
In absence of interactions, $\tilde\epsilon_k=\epsilon_k$, then
$0<\tilde\epsilon_k\ll g(N-1)$ and we have:
$$k\approx g(N-1).\eqno(9)$$
Therefore, supposing a sufficiently large number of interacting
universes, we see that the solutions have necessarily large quantum
numbers and (4) can actually describe our (classical) universe without
introducing other fields or the cosmological constant: the third
quantization mechanism depends only on the gravitational part of the WDW
equation. In the toy model described above, the main result (8) is
strictly dependent on the potential $v(a)$ chosen. This problem can be
overcome by assuming a condensate of baby universes in the ground state
(the foam) and calculating the vertex between an excited universe and the
foam in the presence of an external potential. The condensate provides
the ``energy reservoir'' to ``increase'' the excited universe; moreover,
the presence of the condensate enhances the baby-large universe
interactions with respect to the large-large universe interactions. With
this procedure, a wide class of potentials produces analogous results to
the Hartree--Fock theory discussed previously.

We do not expect the model discussed above to really explain the
properties of our universe, but we hope that this paper will stimulate
work in this direction.

I am very grateful to my friends Prof. Vittorio de Alfaro and Prof. Wanda
Alberico for interesting discussions. I also thank Prof. Kazuo Ghoroku
for important suggestions. Correspondence with Prof. Yasusada Nambu is
gratefully acknowledged.
\beginref
\ref [1] A. Vilenkin, \PR {\bf D37}, 888 (1988).

\ref [2] B. DeWitt, \PR {\bf D160}, 1113 (1967).

\ref [3] J.A. Wheeler, in {\tscors Battelle Rencontres} eds. C.
de Witt and J.A. Wheeler, (Benjamin, 1968).

\ref [4] S.B. Giddings and A. Strominger, \NP {\bf B321}, 481.

\ref [5] K. Ghoroku, \CQG {\bf 8}, 447-452 (1991).

\ref [6] J.B. Hartle and S.W. Hawking, \PR {\bf D28}, 2960 (1983).

\ref [7] O. Bertolami and J.M. Mour\~ao, \CQG {\bf 8}, 1271 (1991).

\ref [8] M. Cavagli\`a and V. de Alfaro, \MPL {\bf A9}, 569
(1994).

\ref [9] A.L. Fetter and J.D. Walecka, {\tscors Quantum Theory of
Many-Particle Systems} (Mc Graw-Hill Inc., 1971).

\endref
\vfill\eject
\bye